# Double resonance Raman study of disorder in CVD-grown single-walled carbon nanotubes

*Rahul Rao,[1*] Jason Reppert,[2] Xianfeng Zhang,[3] Ramakrishna Podila,[2] Apparao M. Rao,[2] Saikat Talapatra,[3] Benji Maruyama[1]*

[1] Air Force Research Laboratory, Materials & Manufacturing Directorate, RXBN, Wright-Patterson AFB, OH 45433

[2]Department of Physics and Astronomy, and COMSET, Clemson University, Clemson, SC, 29634

[3]Department of Physics, Southern Illinois University Carbondale, IL, 62901

*Corresponding Author: Rahul Rao, email: rahul.rao@wpafb.af.mil, Ph: 937-904-4922

**Abstract**

Single-walled carbon nanotubes (SWNTs) with varying degrees of disorder were investigated using multiple-excitation Raman spectroscopy. The lattice disorder was imparted into the nanotubes by the addition of varying amounts of sulfur to the iron catalyst in a thermal chemical vapor deposition process. Changes in the intensities of peaks occurring due to a double resonance Raman process were studied. The intensity of the disorder-induced D band increased with a decrease in the sulfur content. Upon post-synthesis heat treatment, the double resonance process got quenched due to defect healing. The second order G' band and iTOLA bands exhibited a two-peak structure, of which one of the peaks is relatively more sensitive to defects and decreased in intensity with heat treatment.

## 1. Introduction

Raman spectroscopy has evolved into a powerful technique for characterizing carbon-based nanomaterials. Fullerenes, single-walled carbon nanotubes (SWNTs), multi-walled carbon nanotubes (MWNTs), and graphene exhibit distinct Raman features due to their different morphologies thus making it possible to easily distinguish between the various materials [1, 2]. Among these nanoscale forms of carbon, the Raman spectrum of a SWNT is perhaps the richest due to its unique 1D nature [3]. A diameter selective resonance Raman enhancement enables one to characterize individual SWNTs that exhibit low frequency radial breathing modes (RBMs) whose frequencies are inversely proportional to their diameters [2]. Moreover, the tangential vibrational modes have features that are unique to metallic and semi-conducting SWNTs and enable Raman studies sensitive to the electronic structure of the nanotubes [4]. In recent years, Raman studies on individual SWNTs (both isolated and suspended) have revealed great insight into their unique electronic, thermal, and optical properties [5-8].

In typical bulk synthesis methods such as chemical vapor deposition (CVD), the nanotubes produced are in the form of bundles. Such samples contain pristine as well as defective/disordered SWNTs along with impurities such as amorphous carbon and catalyst particles encapsulated in graphitic shells. Defects in the SWNT can be many types including structural defects and amorphous carbon coating on the SWNTs. A common technique for evaluating the degree of defects/disorder in such samples via Raman spectroscopy is to compare the ratio of intensities between the defect-induced D band and the graphitic G band. The exact nature of the D band had been misunderstood for a long time until first Baranov *et al.*, [5] and later Thomsen *et al.* [6] explained the appearance of this feature in graphite due to a double resonance process that involves K point phonons in the Brillouin zone of graphite. Specifically, it was shown that the D band originates from inelastic scattering of a photon with a phonon and elastic scattering by a defect. This scattering process takes place due to real electronic transitions around the K point of the graphite Brillouin zone. Since then, double resonance theory has been applied by Saito *et al*. to explain the occurrence of several overtone and second order features in graphite and carbon nanotubes [7, 8]. One such second order peak is the overtone of the D band that has historically been called the G' band because it is the second most intense feature in the Raman spectrum of graphite. The G' band is a symmetry-allowed second order peak that arises from inelastic scattering by two phonons near the K point of the graphite Brillouin zone [4, 9]. This peak in the Raman spectrum of graphene has had much attention lately due to its distinct lineshape that changes with the number of layers of graphene in a sample [10]. Furthermore, the G' peak has been shown to exhibit sensitivity to doping in SWNTs [11]. In addition to the increase in intensity of the D band, disorder in carbon nanotubes can be observed from changes in the peak linewidths, [12] shifts in peak positions, and changes in intensities of other peaks in the Raman spectra. For example, Fantini *et al*. have observed changes in the intensities of a combination mode at ~2900 $cm^{-1}$ due to functionalization in SWNTs [13]. In another study, Ellis reported an increase in intensity of the higher frequency LOLA peak compared to the iTOLA peak for acid treated double-walled nanotubes (DWNTs) [14]. The LOLA and iTOLA peaks occur due to combinations between the LO and iTO phonons with LA phonons respectively [15].

In this paper we examine some of the above mentioned double resonance features in the Raman spectra of CVD-grown SWNTs where sulfur was used in varying ratios with iron catalyst to impart disorder into the system. We show that in addition to the more common $I_D/I_G$ ratio, one can track changes in other peaks and obtain complementary information that might provide greater insights into the nature of the SWNT sample.

## 2. Experimental Details

A detailed report on the synthesis of the SWNTs has been published previously [16]. Briefly, ferrocene and sulfur powder were mixed in various ratios and placed in a quartz tube furnace (away from the center zone) that was set to a temperature of 1150 ºC. When the temperature of the quartz tube in the region containing ferrocene/sulfur mixture reached ~ 350 ˚C – 400 ˚C, the mixture started to sublimate. At this stage a mixture of argon (Ar) and hydrogen ($H_2$) (Ar 85%) gas was introduced at atmospheric pressure into the quartz tube at a rate of 2000 sccm to carry the vapors into the hot zone of the furnace. The SWNTs were produced in the vapor phase and deposited as bundled films on a stainless steel mesh placed at the end of the quartz tube. For this study, various ratios of iron to sulfur in the precursor were used to change the amount of disorder in the system. Five samples were used having Fe/S powder weight ratios measuring 20/1, 10/1, 5/1, and 3/1, hereafter called sample 20, 10, 5, and 3 respectively. A control sample without any sulfur (sample 0) was also grown for comparison.

Micro-Raman spectra were collected from all samples with a Renishaw inVia Raman microscope coupled with laser excitations $E_{laser}$ = 1.49, 1.96, and 2.33 eV. In addition, FT-Raman spectra ($E_{laser}$ = 1.16 eV) were collected using a Bruker IFS 66v/S Raman spectrometer. The incident laser beam was focused by a 50x objective and the laser power on the samples was kept to a minimum to avoid heating. Whenever possible, spectra were collected from at least a dozen different spots from each sample for in-depth spectral analysis. All the Raman spectra were normalized with respect to the G band intensity and were baseline corrected prior to Lorentzian lineshape analysis.

## 3. Results and Discussion

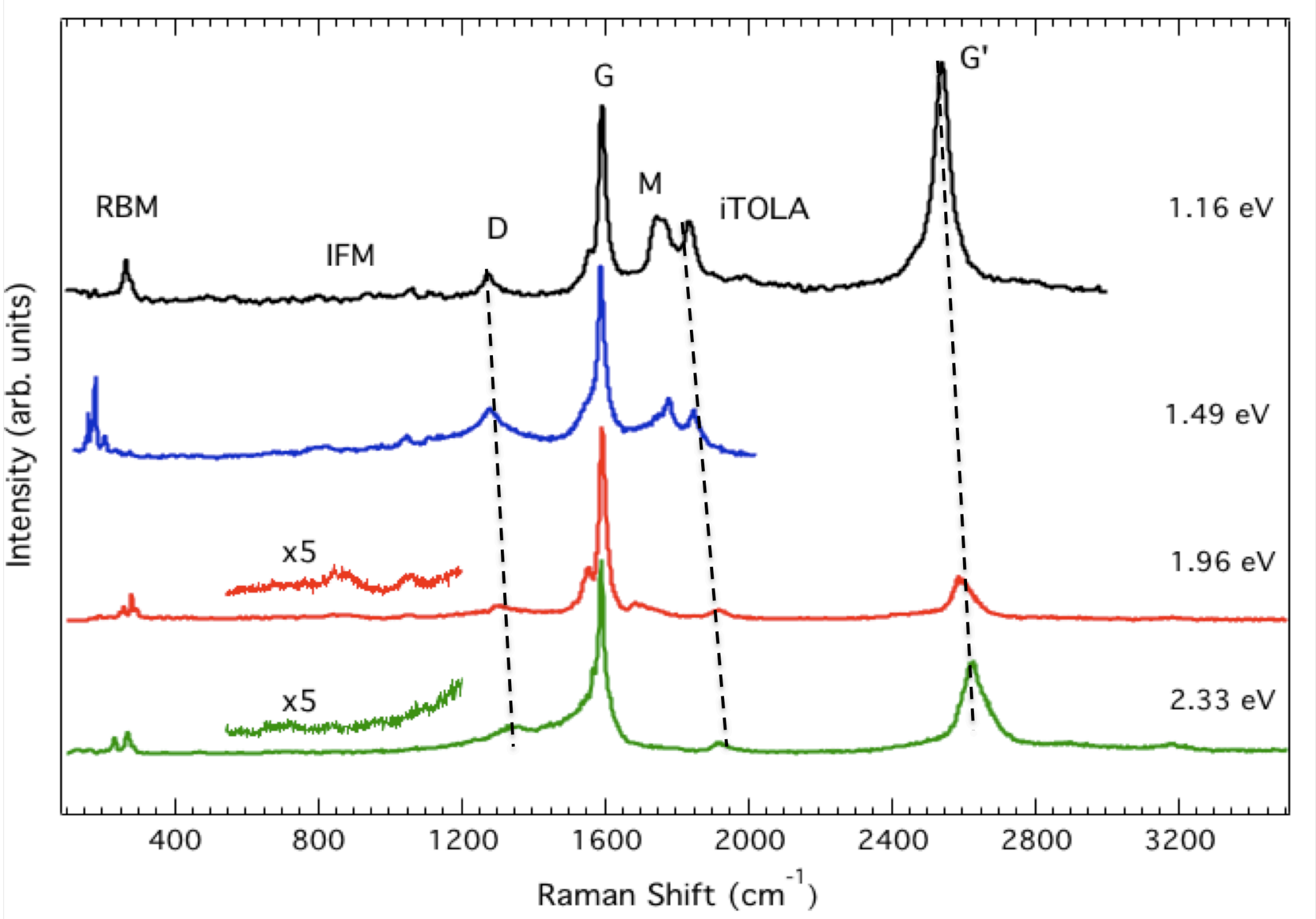

**Figure 1**. Raman spectra of sample 10 (Fe/S ratio: 10/1) collected using various laser excitations.

Figure 1 shows representative Raman spectra from sample 10 collected using different laser excitations. All the Raman spectra in Fig. 1 were normalized with respect to the G band. Due to limitations with the instrumentation, the spectral range for $E_{laser}$ =1.49 eV was limited to 100 – 2000 cm$^{-1}$. The first order RBMs, disorder-induced D band and tangential G bands are marked in the figure. The dispersion of the D band position with laser energy (~ 54 cm$^{-1}$/eV) is shown by the dotted line. In addition to these peaks, two peaks between 1700 and 1900 cm$^{-1}$ are visible in the spectra. The feature near 1750 cm$^{-1}$ has been assigned as an overtone of the out of plane (oTO) infra-red active mode in graphite, which appears at ~860 cm$^{-1}$ in graphite, and is sometimes called the M or 2M band [15]. The feature near 1950 $cm^{-1}$ has been assigned a combination of a phonon from the in-plane transverse optical (iTO) or longitudinal optical (LO) branch and a longitudinal acoustic (LA) branch phonon and is called the iTOLA band [15]. As can be from the dotted line in Fig. 1, the iTOLA band is highly dispersive (~190 $cm^{-1}$/eV) and shifts from ~1830 $cm^{-1}$ to ~1920 $cm^{-1}$ when $E_{laser}$ varies from 1.16 to 2.33 eV [15]. An interesting feature observed in Fig. 1 is the increase in intensity of the M and iTOLA bands with respect to the G band at lower excitation energies. Also observed in Fig. 1 are weak intensity intermediate frequency modes [17] (IFM) between 600 -1200 $cm^{-1}$ (magnified for $E_{laser}$ = 1.96 eV and 2.33 eV) and the dispersive (~106 $cm^{-1}$/eV) second order G' band between 2400 and 2800 $cm^{-1}$.

The intensity of the G' band appeared to change with respect to the G band for various excitations. In particular, the G' band intensity was quite high for samples excited by $E_{laser}$ = 1.16 eV. Such an increase in intensity of the G' band with laser energy has been observed previously in nanocrystalline graphite [18]. However, in graphitic samples there is a concomitant increase in intensity of the D band with decreasing laser energy, [18, 19] which is not very obvious in the Raman spectra shown in Fig. 1. On the other hand, for SWNTs, the D and G' band intensities in metallic samples are typically higher than semiconducting nanotubes due to a larger electron-phonon matrix element for the TO phonon mode at the K point in the Brillouin zone [20-22]. A comparison of the RBM peaks in the present data with the so-called Kataura plot [1] in shown in Fig. 2 shows that semiconducting SWNTs are predominantly excited by $E_{laser}$ = 1.49 eV and 1.16 eV, while metallic SWNTs are the predominant species excited by $E_{laser}$ = 2.33 eV. This might explain the larger intensity of the G' band and asymmetric G band lineshape in the SWNTs excited by $E_{laser}$ = 2.33 eV compared to $E_{laser}$ = 1.96 eV (see Fig. 1). In all the spectra collected from the samples used in this study, the intensities of the G' band were observed to be consistently higher when measured with $E_{laser}$ = 1.16 eV. In addition the intensities of the M and iTOLA band were also higher for $E_{laser}$ = 1.49 eV and 1.16 =/eV, thus suggesting a unique dependence of the two phonon double resonance process on laser energy, the origin of which could be related to the phonons involved in the double resonance process related to the D and G' bands. The D band is typically attributed to the inelastic scattering of a π electron in the conduction band by a TO phonon, [23] which implies that the G' band occurs due to scattering by two TO phonons. However, Raman intensity calculations by Kim *et al*. revealed that the G' band intensity is stronger for semiconducting SWNTs compared to metallic SWNTs for $E_{laser}$ < 1.5 eV if the G' band is assumed to occur due to scattering by two LO phonons instead of TO phonons [22]. More studies are needed in this regard to determine the type and nature of phonons involved in the double resonance scattering process related to the D and G' band.

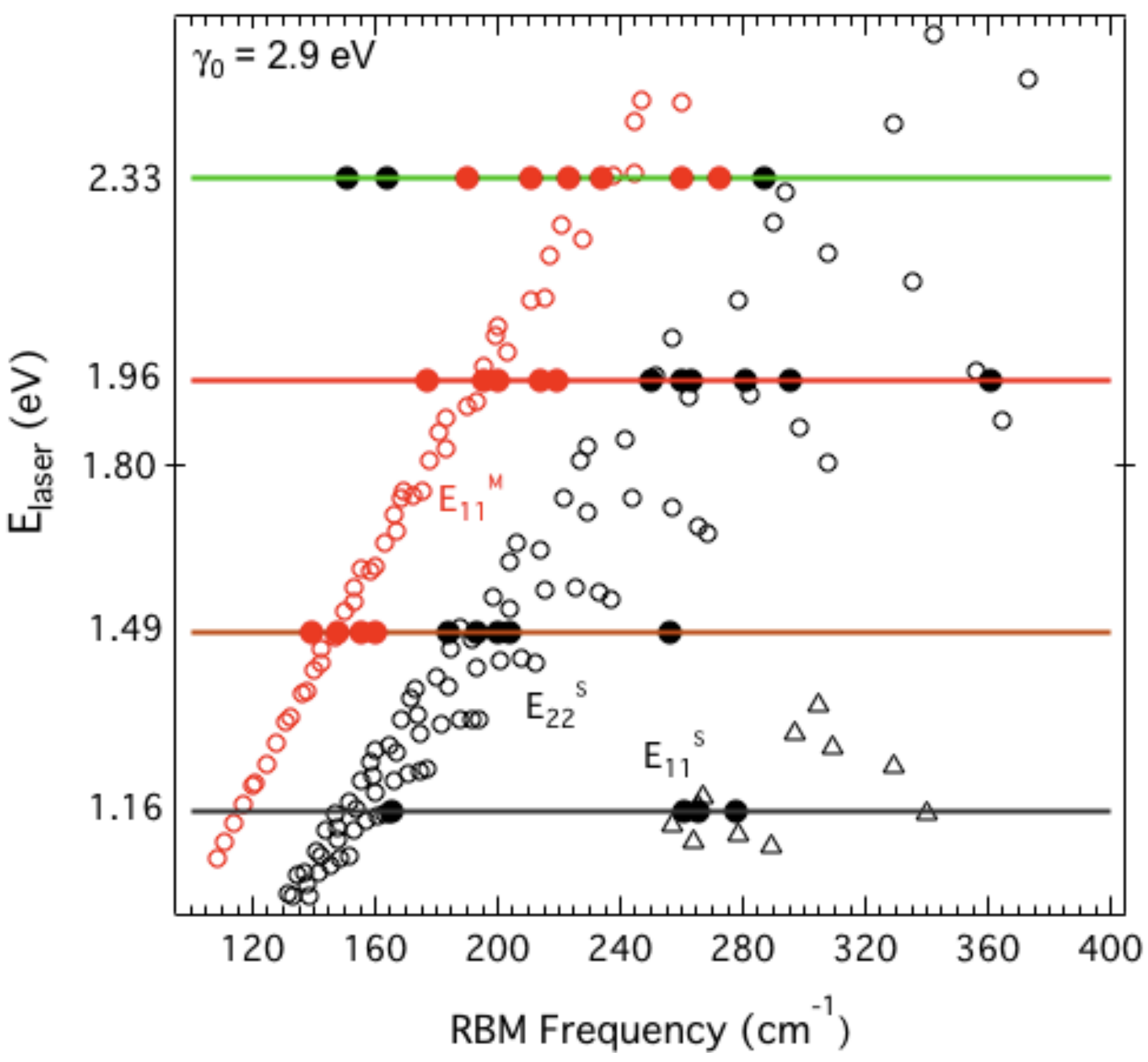

**Figure 2**. Experimental Kataura plot showing the first ($E_{11}^{S}$) and second ($E_{22}^{S}$) optical transitions of semiconducting, and first ($E_{11}^{M}$) transition of metallic nanotubes. The data points for $E_{22}^{S}$ (open circles) and $E_{11}^{S}$ (open triangles) have been taken from Ref. [24] and [25] respectively, and the data for the metallic tubes (open circles) has been taken from Ref. [26]. RBM frequencies measured in the present study are plotted (filled circles) along with horizontal lines indicating the various laser excitation energies.

The D band is a defect induced double resonance peak that occurs due to inelastic scattering by a phonon and elastic scattering by a defect [10,11,13]. Hence disordered samples should exhibit higher D band intensities and the ratio of integrated intensity of the D band to that of the G band ($I_D/I_G$) is a good indicator of sample quality. Figure 3 shows the $I_D/I_G$ ratios (left y-axis) for samples 20, 10, 5 and 3. The sample synthesized without any sulfur (sample 0) exhibited the highest $I_D/I_G$ ratio among all samples [16]. In addition, the yield of SWNTs in sample 0 was poor, hence the spectra from sample 0 will not be discussed further. Samples 10 and 5 exhibit the lowest $I_D/I_G$ ratio and hence highest quality among all samples. Independent observations by electron microscopy studies have proven the high quality of samples 10 and 5 [16]. The linewidths of the G and D bands ranged from ~10-30 $cm^{-1}$, and ~25-60 $cm^{-1}$, respectively. Similar values for the D band linewidth for SWNT bundles have been reported in the literature which strongly suggests that the D band observed in these samples is intrinsic to SWNTs rather than amorphous carbon, whose linewidths are typically higher than those for SWNTs [12, 27]. Moreover, the D band frequencies and their laser energy dependence measured in this study are closer to that reported for SWNTs, rather than other forms of $sp^2$ carbon which typically exhibit higher frequency D bands [12]. The increase in D band intensity for samples 20 and 3 compared to the other samples are presumably due to changes in the nanotube structure brought upon by interaction of sulfur with the iron and carbon during growth. A detailed study on the diameter and structural changes of the nanotubes due to varying amounts of sulfur with respect to iron will be published elsewhere. For the present study, it can be assumed that the various amounts of sulfur in the precursor created varying amounts of structural disorder in the SWNTs that can be distinguished by analyzing their Raman spectra.

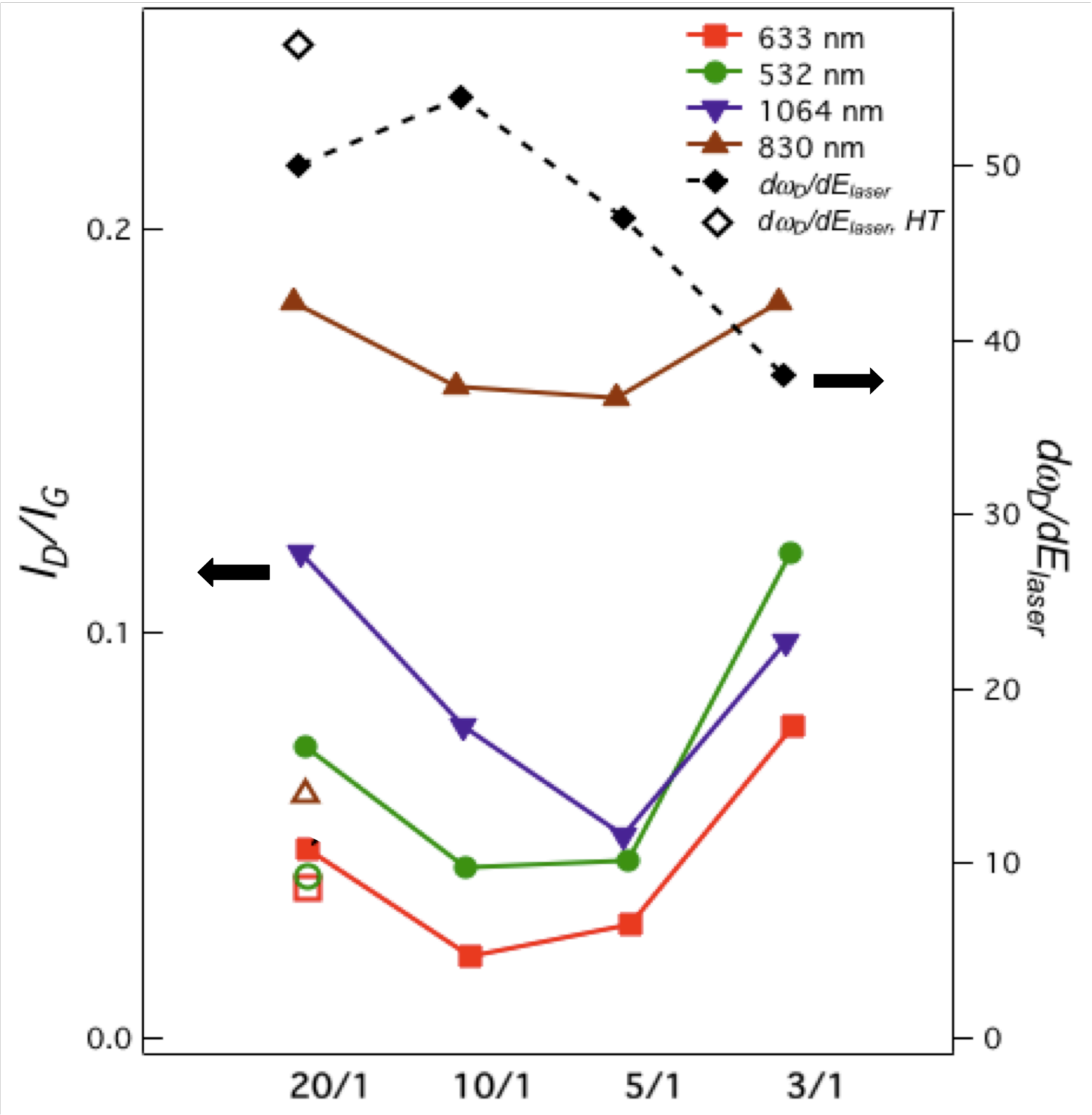

**Figure 3**. $I_D/I_G$ ratios (left axis) for the four samples containing varying amounts of iron and sulfur, and slope of D band frequency versus $E_{\text{laser}}$ ($d\omega/dE_{\text{laser}}$, right axis). The filled (open) symbols correspond to the $I_D/I_G$ & $d\omega/dE_{\text{laser}}$ data before (after) the heat treatment.

Also plotted in Fig. 3 (right y-axis) is the slope of the dispersion of the D band ($d\omega_{\text{D}}/dE_{\text{laser}}$) exhibited by the four samples. The typical dispersion of the D band for SWNT bundles has been reported to be ~50 $\text{cm}^{-1}$/eV [9, 28], although Kukovecz *et al*. reported a decrease in the slope of the D band dispersion and upshift in D band frequency after chemical treatment [29]. From Fig. 3, it can be seen that $d\omega_{\text{D}}/dE_{\text{laser}}$ follows the $I_D/I_G$ ratio where the highest (lowest) slope is associated with the lowest (highest) $I_D/I_G$ ratio. To investigate changes in $d\omega_D/dE_{laser}$ and the $I_D/I_G$ ratio, sample 20 was heat-treated (HT) in the presence of argon gas at 1000 °C. Although the $I_D/I_G$ ratio was higher for sample 3, sample 20 was chosen for the heat treatment study because it contained a majority of SWNTs compared to sample 3, which contained MWNTs along with SWNTs [16]. The slope of the D band dispersion and $I_D/I_G$ ratios from sample 20 after heat treatment collected using $E_{laser}$ = 1.49, 1.96, and 2.33 eV are plotted in Fig. 3 (open symbols). A decrease in the $I_D/I_G$ ratio and increase in $d\omega_D/dE_{laser}$ for sample 20 after heat treatment is indicative of defect healing during high temperature annealing. A slight downshift (~4 $cm^{-1}$) in D band frequency was also observed in the heat-treated sample. It is interesting to note that the $I_D/I_G$ ratios for the non-HT have different values for different laser excitations, while the $I_D/I_G$ ratios for the heat-treated samples are very similar. Variation of the D band intensity has been observed previously for certain laser excitation energies and was attributed to resonance with metallic SWNTs as mentioned in the previous section [20, 21]. In addition, a dependence of the $I_D/I_G$ ratio on laser energy (inversely proportional to $E^4_{laser}$) has also been observed for nanocrystalline graphite [30] and for SWNTs with different lengths [20]. Such a dependence of the $I_D/I_G$ ratio on $E_{laser}$ was also confirmed for the samples measured in this study. Interestingly however, it appears that the double resonance process gets quenched as the defects are annealed away in the SWNTs, leading to a loss in dependency of the $I_D/I_G$ ratio on laser energy in the annealed SWNTs. This suggests that the dispersion of the D band position might be a better indicator of disorder compared to the intensity of the D band with respect to the G band.

Figure 4 shows Raman spectra collected from sample 20 in the second order G' band region. Since the G' band also originates due to a double resonance process, it can be used to study the electronic structure of graphitic materials and changes caused by defects or doping [11, 31, 32]. In the past, a two-peak structure has been observed in the G' band region for graphite and was ascribed to the 2D stacking of graphene layers [9]. In SWNTs, a two peak G' structure was observed both in isolated and bundled samples and was attributed to differences in resonance with the incident and scattered light [33, 34]. A two-peak structure has also been observed in double-walled nanotube (DWNT) samples and the lower (higher) frequency peaks have been attributed to arising from the inner (outer) tubes respectively [35, 36]. More recently, Maciel *et al.* observed a two-peak structure in SWNTs doped with nitrogen, boron, and phosphorus. They attributed one of the peaks to a defect related peak caused by the dopant and its position changed relative to the other (pristine) peak depending on the type of dopant, i.e. electron donor or acceptor [11]. In a different study, Maciel *et al.* also observed a two peak structure in the G' band of pristine SWNTs [37]. They attributed the lower frequency G' peak to defects in the samples and reported a decrease in the intensity of the lower frequency peak upon heat treatment at 400 ºC [37]. A similar trend can be seen from the spectra collected from sample 20 (see Figs. 4(a) and 4(b)) where the G' band is composed of two peaks and the intensity of the lower peak ($G'_l$) decreases relative to the intensity of the higher peak ($G'_h$) after heat treatment in argon environment at 1000 °C. In the spectra collected with $E_{laser}$ = 1.96 eV, a third peak can be observed at ~2450 $cm^{-1}$ (Fig. 4a) and has been attributed to a non-dispersive overtone of a LO phonon around the K point in the Brillouin zone of graphite [38]. No significant change in the position of the peaks was observed due to increasing sulfur content, suggesting that the amounts of sulfur in the samples were too low to exhibit the kind of shifts seen due to substitutional doping in SWNTs by nitrogen or boron as described in Ref. 11.

A plot of the ratio of integrated intensities between the lower and higher frequency G' peaks for the four different samples is shown in Fig. 4(c). Sample 5 exhibits the lowest intensity of the G'$_l$ peak relative to the G'$_h$ peak followed by sample 10. This trend is similar to the trend shown by the $I_D/I_G$ ratios plotted in Fig. 3, confirming that the lower frequency G' peak arises due to defects present in the sample. Moreover, the decrease in intensity of the G'$_l$ peak in sample 20 after heat treatment lends support to the above argument. The slope of the dispersion of the G'$_l$ and G'$_h$ peaks plotted in Fig. 5 exhibit a similar trend as the slope of the D band. Interestingly, the slope of the G'$_l$ peak for sample 20 increased dramatically after high temperature heat-treatment, while the slope of the G'$_h$ peak decreased slightly. This observation provides further proof that the G'$_l$ peak is defect-induced in the SWNT samples. The dispersion values for both peaks within the G' band matches the average reported value of ~100 $cm^{-1}$/eV [13,41].

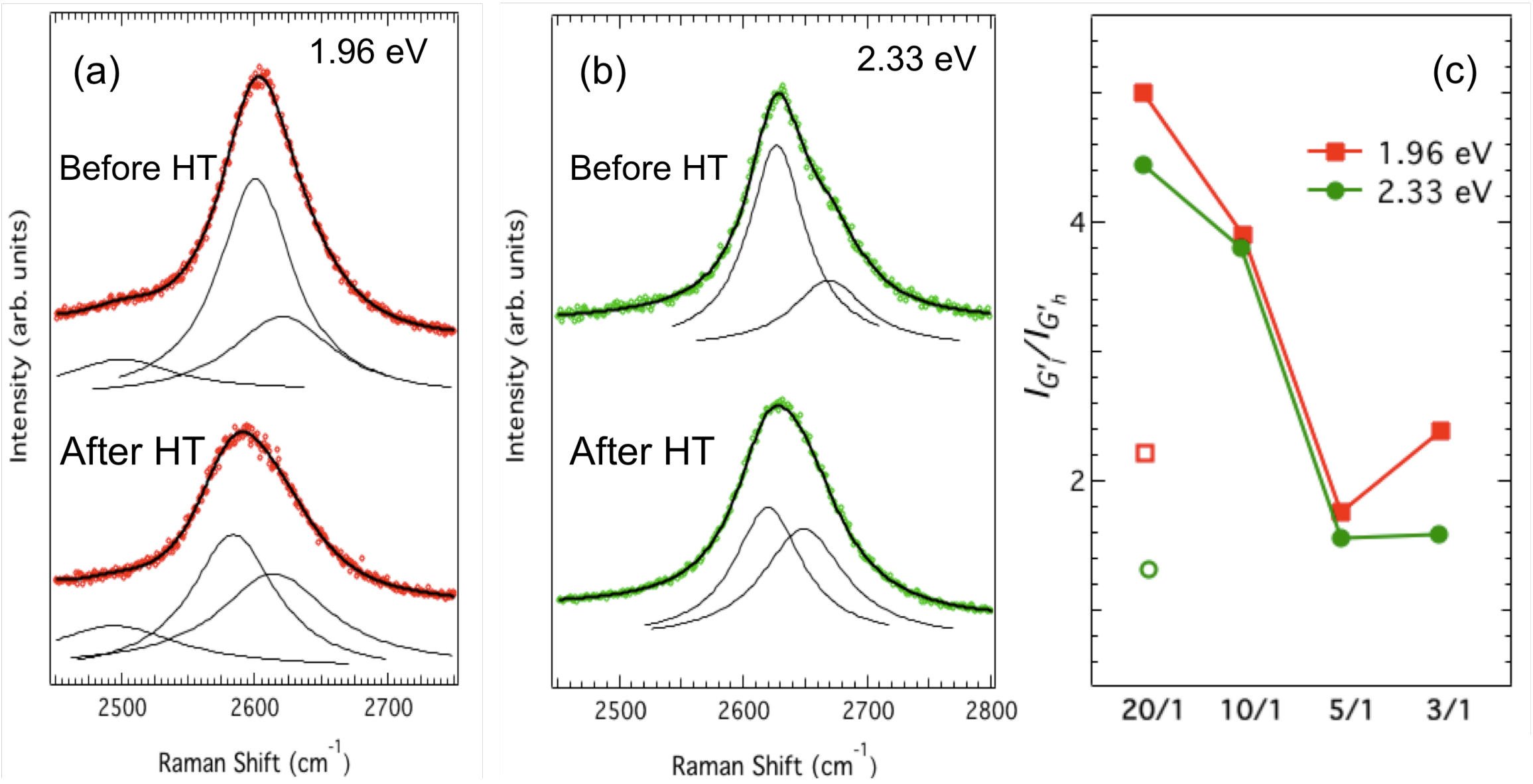

**Figure 4**. Raman spectra from sample 20 in the second order G' band region before and after heat treatment at 1000 °C. Spectra were collected using (a) 1.96 eV, and (b) 2.33 eV laser excitation. (c) Plot of intensity ratio between lower ($G'_l$) and higher ($G'_h$) frequency *G'* band for the pristine (filled symbols) and heat-treated (open symbols) samples used in this study.

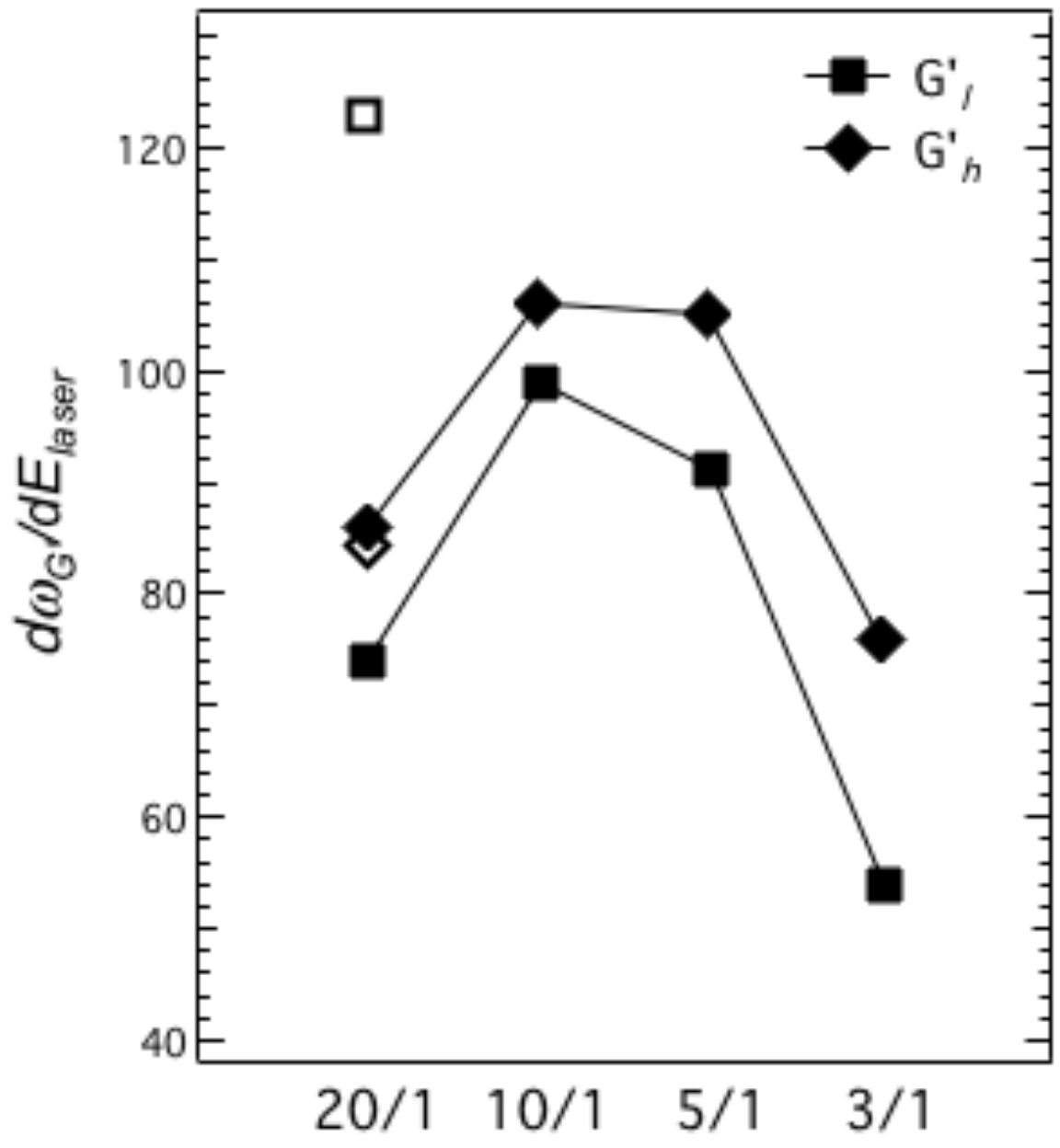

**Figure 5**. Slope of the dispersion of the lower (G'$_l$) and higher (G'$_h$) peaks in the G' band of the pristine (filled symbols) and heat-treated (open symbols) SWNTs used in this study.

In addition to the D and G' bands in the SWNT Raman spectrum, Fantini *et al*. observed changes in other double resonance Raman modes in functionalized SWNTs. Specifically, they reported a decrease in intensity of modes between 600 – 1200 cm$^{-1}$, also called the intermediate frequency modes (IFM) [13]. The Raman spectra obtained from the samples in the present study did not have enough signal-to-noise ratios (see magnified insets in Fig. 1) to distinguish clearly between various peaks in the IFM region. In addition to the IFM, Fantini *et al*. observed an increase in the intensity of a peak at ~2900 cm$^{-1}$ in the Raman spectra of the functionalized SWNTs, which they assigned to a combination mode between the D band and the D' band occurring at ~ 1620 cm$^{-1}$ in disordered SWNTs [13]. The D' band is also a defect-induced peak which originates from an intravalley double resonance process [6]. This peak is observed in the Raman spectra of defective graphite and MWNTs, but is difficult to resolve in the Raman spectra from SWNT bundles due to the presence of the $E_{2g}$ peak occurring at ~1605 $cm^{-1}$. The D' peak has however, been observed previously in isolated nitrogen-doped SWNTs and could be related to defects intrinsic to the structure of the SWNTs [31, 32]. In this study, no obvious D' peak at ~1620 $cm^{-1}$ was observed in any of the spectra collected from the various samples. A peak at ~2900 $cm^{-1}$ was indeed observed as a weak feature in some of the Raman spectra in the present study, although no noticeable trend could be discerned with respect to change in iron/sulfur ratio.

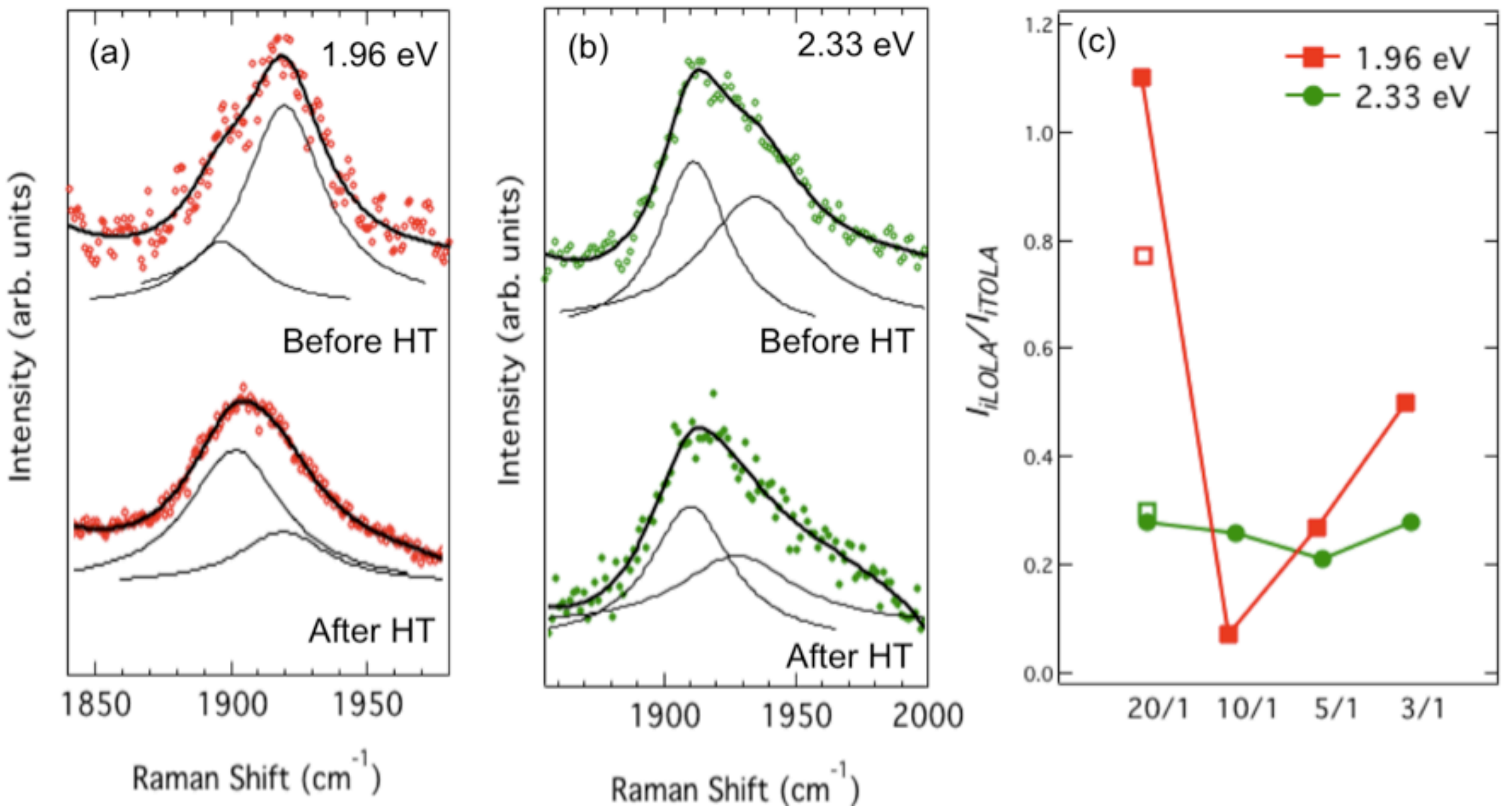

**Figure 6**. Raman peaks in the iTOLA region of sample 20 collected with (a) $E_{laser}$ = 1.96 eV, and (b) $E_{laser}$ = 2.33 eV before and after heat-treatment, (c) Plot of intensity ratio between lower (*LOLA*) and higher (*iTOLA*) frequency band for the pristine (filled symbols) and heat-treated (open symbols) samples used in this study.

Finally we turn our attention to a weak intensity feature called the iTOLA band which appears at ~1900 $cm^{-1}$ when excited with visible excitation. As mentioned above, this band has been assigned to a combination of a phonon from the iTO or LO branch and a LA phonon [11,18,19]. In addition, the band can often be resolved into two peaks when visible excitation is used, and the low and high frequency peaks have been assigned as combination modes between iTO and LO phonons with LA phonons respectively [14]. Ellis recently reported an increase in intensity of the higher frequency LOLA peak compared to the iTOLA peak for acid treated DWNTs [14]. The Raman spectra in the iTOLA region from sample 20 are shown in Fig. 6(a) and 6(b). The peaks exhibit a two-peak structure indicating the presence of the lower frequency iTOLA and higher frequency LOLA peak. In addition, the intensity of the LOLA peak decreases relative to the iTOLA peak after heat treatment, indicating that the LOLA peak is more sensitive to defects than the iTOLA peak. The change in intensities of the iTOLA and LOLA peaks are very prominent in the samples excited by $E_{\text{laser}}$ = 1.96 eV (see Fig. 6(a)). Furthermore, a plot of the ratio of intensities of the LOLA peak to iTOLA peak for the various samples reveals a trend that is very similar to trends shown by the $I_D/I_G$ ratios (Fig. 3) as well as the $I_{G'l}/I_{G'h}$ ratios (Fig. 4c).

Although the trends shown by the $I_D/I_{G,}$ $I_{G'l}/I_{G'h}$ and the $I_{LOLA}/I_{iTOLA}$ ratios are very similar, subtle differences between them indicate variability in the type and nature of disorder in the various samples. For example, the $I_{G'l}/I_{G'h}$ ratio is the lowest for sample 5 while the $I_D/I_G$ ratio for the same excitation (1.96 and 2.33 eV) is lowest for sample 10. The reason for this difference is unclear at present and a detailed high-resolution electron microscopy study is underway to determine the changes in microstructure brought about by changing sulfur amounts in samples 5 versus 10[1]. In addition, the trends in the $I_D/I_G$ ratio varies slightly between $E_{\text{laser}}$ = 1.96 eV and 2.33 eV compared to $E_{\text{laser}}$ = 1.16 eV and 1.49 eV. This difference is probably due to the fact that fewer Raman spectra were collected from the samples using the NIR excitations compared to visible excitations. For the particular diameter range of SWNTs used in this study, there were fewer nanotubes resonant with the $E_{\text{laser}}$ = 1.16 eV and 1.49 eV compared to $E_{\text{laser}}$ = 1.96 eV and 2.33 eV, as can be seen in the Kataura plot shown in Fig. 2. Finally, the differences in trends described above could also be caused by diverse defect scattering mechanisms due to TO or LO phonons. More experimental and theoretical effort is needed to understand the role of defects in the double resonance scattering processes in carbon nanotubes.

## 4. Conclusions

The addition of sulfur to iron catalyst leads to structural changes and various degrees of disorder in thermal CVD-grown SWNTs. The degree of disorder changes with varying iron to sulfur ratio and our results show that such changes can be observed by analyzing various peaks in the Raman spectra of the SWNTs. In this study, five samples having various degrees of disorder were analyzed via Raman scattering using $E_{\text{laser}}$ = 1.16, 1.49, 1.96, and 2.33 eV. The $I_D/I_G$ ratios of samples grown with iron to sulfur weight ratios of 10/1 and 5/1 were lower than those of samples 20/1 and 3/1. It was found that the $I_D/I_G$ ratios measured with different laser excitations approach similar values following high temperature heat treatment, indicating a quenching of the double resonance process. In addition to changes in the D band, disorder in the SWNTs was also observed in the two-peak structure of the second order G' band as well as the iTOLA band. Changes in the structure of the G' and iTOLA peaks after heat treatment were concomitant with changes in the $I_D/I_G$ ratio and slope of the D band dispersion. No change in peak position of the G' band was observed with increasing sulfur content, indicating a lack of evidence for substitutional doping. Increased intensities of the G' and iTOLA bands with decreasing laser energy suggests a dependence of the two phonon double resonance process with laser energy. Subtle differences between the various trends reported in Section 3 suggest differences in the nature of the disorder in the samples that could possibly be discerned through high-resolution microscopy studies. Such a study may also shed further light into the physical phenomena that cause the appearance of these disorder-induced peaks in the Raman spectra of SWNTs and other graphitic carbon materials.

[1] Rao R, Pierce N, Zhang X, Talapatra S, Maruyama B. Unpublished results.

**Acknowledgement**

R.R. and B.M. acknowledge support from the AFOSR and the National Research Council associateship program. AMR acknowledges support through US AFOSR Grant Number FA9550-09-1-0384. ST acknowledges the support by the Illinois Department of Commerce and Economic Opportunity through the Office of Coal Development and the Illinois Clean Coal Institute (ICCI grant No. 10/7B-5), and NSF ECCS (grant No. 0925708).

**List of Figure Captions**

**Figure 1**. Raman spectra of sample 10 (Fe/S ratio: 10/1) collected using various laser excitations.

**Figure 2**. Experimental Kataura plot showing the first ($E_{11}^{S}$) and second ($E_{22}^{S}$) optical transitions of semiconducting, and first ($E_{11}^{M}$) transition of metallic nanotubes. The data points for $E_{22}^{S}$ (open circles) and $E_{11}^{S}$ (open triangles) have been taken from Ref. [24] and [25] respectively, and the data for the metallic tubes (open circles) has been taken from Ref. [26]. RBM frequencies measured in the present study are plotted (filled circles) along with horizontal lines indicating the various laser excitation energies.

**Figure 3**. $I_D/I_G$ ratios (left axis) for the four samples containing varying amounts of iron and sulfur, and slope of D band frequency versus $E_{\text{laser}}$ ($d\omega/dE_{\text{laser}}$, right axis). The filled (open) symbols correspond to the $I_D/I_G$ & $d\omega/dE_{\text{laser}}$ data before (after) the heat treatment.

**Figure 4**. Raman spectra from sample 20 in the second order G' band region before and after heat treatment at 1000 °C. Spectra were collected using (a) 1.96 eV, and (b) 2.33 eV laser excitation. (c) Plot of intensity ratio between lower ($G'_l$) and higher ($G'_h$) frequency *G'* band for the pristine (filled symbols) and heat-treated (open symbols) samples used in this study.

**Figure 5**. Slope of the dispersion of the lower ($G'_l$) and higher ($G'_h$) peaks in the G' band of the pristine (filled symbols) and heat-treated (open symbols) SWNTs used in this study.

**Figure 6**. Raman peaks in the iTOLA region of sample 20 collected with (a) $E_{\text{laser}}$ = 1.96 eV, and (b) $E_{\text{laser}}$ = 2.33 eV before and after heat-treatment, (c) Plot of intensity ratio between lower (*LOLA*) and higher (*iTOLA*) frequency band for the pristine (filled symbols) and heat-treated (open symbols) samples used in this study.